\documentclass{elsart}

\usepackage{graphicx}

\usepackage{amssymb}

\def\be{\begin{equation}}
\def\ee{\end{equation}}
\begin{document}
\begin{frontmatter}

\title{Detection of high energy cosmic rays with the resonant gravitational wave detector NAUTILUS and EXPLORER}

\author[label1]{P.~Astone},
\author[label2]{D.~Babusci},
\author[label3,label4]{M.~Bassan},
\author[label1,label5]{P.~Bonifazi},
\author[label8]{G. Cavallari},
\author[label3,label4]{E.~Coccia},
\author[label3]{S.~D'Antonio},
\author[label3,label4]{V.~Fafone},
\author[label2]{G.~Giordano},
\author[label2]{C.~Ligi},
\author[label2]{A.~Marini},
\author[label2]{G.~Mazzitelli},
\author[label6]{Y.~Minenkov},
\author[label6]{I.~ Modena},
\author[label2]{G.~Modestino},
\author[label3,label4]{A.~Moleti},
\author[label1,label7]{G.~V.~Pallottino},
\author[label2,label3]{G.~Pizzella},
\author[label2]{L.~Quintieri},
\author[label3]{A.~Rocchi},
\author[label2]{F.~Ronga\corauthref{fr}},
\author[label3,label5]{R.~Terenzi},
\author[label3,label5]{M.~Visco}

\address[label1] {INFN (Istituto Nazionale di Fisica Nucleare) sezione di Roma I ,I 00185  Roma, Italy}
\address[label2] {INFN  Laboratori Nazionali di Frascati, I 00044 Frascati, Italy}
\address[label3] { INFN Sezione di Roma II, I 00133 Roma, Italy}
\address[label4] {Dipartimento di Fisica, Universit\`a di Roma "Tor Vergata", I 00133 Roma, Italy}
\address[label5] {Istituto di Fisica dello Spazio  Interplanetario INAF, I 00133 Roma, Italy}
\address[label6] {INFN Laboratori Nazionali del Gran Sasso, I 67010 Assergi, Italy}
\address[label7] {Dipartimento di Fisica, Universit\`a di Roma  "La Sapienza", I  00185 Roma, Italy}
\address[label8] {CERN, CH 1211 Geneva, Switzerland}
\corauth[fr]{Corresponding author} 

%\maketitle
\date{}

\begin{abstract}
The cryogenic resonant gravitational wave detectors  NAUTILUS and EXPLORER, made of  an aluminum alloy bar, can detect cosmic ray showers. At temperatures above 1 K, when the material is in the normal conducting state, the measured signals  are in good agreement with the values expected based on the cosmic rays data and  on the thermo-acoustic model. When NAUTILUS was operated at the temperature of 0.14 K, in superconductive state,  large signals produced by cosmic ray interactions, more energetic than expected, were recorded. The NAUTILUS data in this case are in agreement with the measurements done by a dedicated experiment on a particle beam. The biggest recorded event was in EXPLORER and excited the first longitudinal mode to a vibrational energy of $\sim670$ K,  corresponding to $\sim360$ TeV absorbed in the bar. Cosmic rays can be an important background in future acoustic detectors of improved sensitivity. At present, they represent a useful tool to verify the gravitational wave antenna performance.
\end{abstract}

\end{frontmatter}

\section{Introduction}

Cosmic ray showers can excite mechanical vibrations in a metallic cylinder at its resonance frequencies and can provide an accidental background for experiments searching gravitational waves (gw): this possibility 
was suggested many years ago and a first search, ending with a null result, was carried out with room temperature Weber type resonant bar detectors \cite{Ezrow:1970yg}.
  
More recently,  the cryogenic resonant gw detector  NAUTILUS has been equipped with shower detectors and the interaction of cosmic ray with the antenna has been studied in detail.

The first detection  of cosmic ray signals in a gw detector took place in 1998 in NAUTILUS. During this run many events of very large amplitude were detected. This  unexpected result suggested  in 2002 the construction of a cosmic ray detector even for the EXPLORER detector.

In section \ref{TAM} we briefly recall the main features of the Thermo-Acoustic Model (TAM), that successfully describes the interaction between a solid elastic resonator and  a charged particle, or a beam of such particles; some of these features are extended to the regime of superconducting metal for the elastic resonator.

In section \ref{cosm_descr} we describe the  NAUTILUS and EXPLORER cosmic ray detectors and we specialize the TAM model to the interaction with cosmic rays computing the expected event rates.

In section \ref{antennas} we describe the results of coincidence measurements between the output of each gw antenna and its respective cosmic ray monitor, in different periods of data taking:  for NAUTILUS during the year 1998, with the antenna in superconductive state, and then in the years from 2003 to 2006, while for EXPLORER in the period  from 2003 to 2006.  The data are interpreted  with the help of some results obtained by the RAP experiment. Finally, some conclusions of this extended analysis are drawn: we show the good agreement of our data with the TAM predictions and  the consistency of data taken by two detectors with various different experimental setups (temperature, bandwidth, readout and acquisition hardware and software).
As a central result, the relevance of the conducting state of the antenna material on the strength of the interaction is proven.

We have then learned that cosmic rays can also provide a good calibration source for resonant gw detectors, as they very closely mimic the signal expected by  short bursts of gravitational waves, i.e. the tidal excitation of the longitudinal modes.

\section{The thermo-acoustic model and its experimental validation with particle beams}\label{TAM}
 
The interaction of energetic charged particles with a normal mode of an extended elastic cylinder has been extensively studied over the years, both on the theoretical and on the experimental aspect.

The first experiments aiming to detect mechanical oscillations in metallic targets due to impinging elementary particles were carried out by Beron and Hofstander as early as in 1969 \cite{beron1,beron2}.
A few years later, Strini et al. \cite{grassi} carried out an experiment with a small metallic cylinder and measured the cylinder oscillations.  The authors compared the data against the TAM (Thermo Acoustic Model) in which the longitudinal vibrations are originated from the local thermal expansion caused by the warming up due to the energy lost by the particles crossing the material. In particular, the vibration amplitude is directly proportional to the ratio of two thermophysical parameters of the material,  namely the thermal expansion coefficient and the specific heat at constant volume. The ratio of these two quantities appears in the definition of  the Gr\"{u}neisen parameter $\gamma$. It turns out that while the two thermophysical parameters vary with temperature, $\gamma$ practically does not, provided the temperature is above the material superconducting $(s)$ state  critical temperature.

Detailed calculations,  successively refined by several authors \cite{allega,deru,deru1,amaldi,liu} agree in 
predicting, for the excitation energy  $E$ of the fundamental vibrational mode of an aluminum cylindrical bar, the following equation: 
 \be
 E=\frac{4}{9\pi}\frac{\gamma^2}{\rho L v^2}(\frac{dW}{dx})^2[sin(\frac{\pi z_o}{L})\frac{sin[(\pi l_ocos(\theta_o)/2L]}{\pi Rcos(\theta_o)/L}]^2
 \label{eliub}
 \ee
where  $L$ is the bar length, $R$ the bar radius, $l_o$ the length of the particle track inside the bar, $z_o$ the distance of the track mid point  from one end of the bar, $\theta_o$ the angle between the particle track and the axis of the bar, $\frac{dW}{dx}$ the energy loss of the particle in the bar, $\rho$ the density, $v$ the  longitudinal  sound velocity in the material .
This relation  is valid for the material normal-conducting $(n)$ state  and some authors (see ref.  \cite{allega,deru}) have extended the model to a super-conducting  ($s$) resonator, according to a scenario in which the vibration amplitude is due to two pressure sources, one due to $s-n$ transitions in small regions centered around the interacting particle tracks and the other due to thermal effects in these regions now in the $n$ state. It is important to note, at this point, that  a gw bar antenna, used as particle detector, has characteristics very different from the usual particle detectors which are sensitive only to ionization losses \cite{deru1}\cite{Astone:1992zf}:  indeed an acoustic resonator can be seen as a zero threshold calorimeter, sensitive to a vast range of energy loss processes.
\\

As anticipated in the introduction, the first detection of signals in a gw detector output due to cosmic ray events, took place in  1998.  The NAUTILUS detector, a bar made of the aluminum alloy Al 5056   was operated at a thermodynamic temperature $T=0.14$ K~\cite{cosmico1},  i.e. below the $s$ transition temperature $T_c \simeq0.9 K$. During this run, many  events of unexpectedly large amplitude were detected. This result suggested an anomaly either in the model or in the cosmic ray interactions\cite{cosmico2}. However the observation was not confirmed in the 2001 run with NAUTILUS at $T=1.5$ K~\cite{cosmico3} and therefore we made the hypothesis that the unexpected behavior be due to the superconducting state of the material.
An experiment (RAP) \cite{rap} was then planned at the INFN Frascati National Laboratory to study the vibration amplitude of a small Al 5056 bar caused by the hits  of a 510 MeV electron beam. The experiment was also motivated by the lack of complete knowledge of the thermophysical parameters  of the alloy Al 5056 at low and ultra-low temperatures. We summarize here the main results obtained by the experiment:
 
(a ) in Al 5056 at $T \sim 4K$~\cite{rap} RAP measured a ratio, $\alpha_{n}=1.15$,  between the measured vibration amplitudes and the expectations based on the thermo-acoustic model of eq.\ref{eliub} using the TAM parameters known for  pure aluminum;

(b) the experimental verification, made on a pure niobium bar, that the amplitude depends on the material conduction state~\cite{rap1} and

(c) for Al 5056 at $T\sim0.5$ K  a ratio, $\alpha_{s}=3.7$,  between the measured amplitudes at $T\sim 0.5 K$ and $T \sim 1.5 K$~\cite{rap2} has been measured. 

While $\alpha_{n}$ can be considered a small correction due to our inexact knowledge of relevant thermophysical parameters of our material,  a value of $\alpha_{s}$ so different from unity indicates that more complex interactions, as mentioned above, take place in the superconducting alloy.

\section{The  cosmic rays detectors  of NAUTILUS and EXPLORER, and their expected rates}\label{cosm_descr}

The gw  detector NAUTILUS\cite{Astone:1997gi} is located in Frascati (Italy) National Laboratories of INFN,  at about 200 meters above sea level. It is equipped
 with a cosmic ray detection telescope made of seven
layers of gas detectors (streamer tubes) for a total of 116 counters \cite{coccia}.
 Three superimposed layers, each with an area of $36~m^2$, 
are located above the cryostat. Four superimposed layers
are below the cryostat, each with area of $16.5~m^2$.
The signal from each counter is digitized to measure the charge  that
 is proportional to the number of particles. The detector is capable of  measuring particle densities up to $1000~ \frac{particles}{m^2}$
without large saturation effects. During normal runs only showers are detected, with a typical threshold of the order of  $2 ~\frac{particles}{m^2}$ in the lower detectors. 

Single particles, cosmic rays muons, are collected every day in short runs, with the aim of calibrating the detectors.  The systematic error on the absolute number of  particles crossing the apparatus is of the order of 25\%. The cosmic ray and the antenna data acquisition systems are independent. A  GPS clock is used to synchronize the antenna and cosmic ray acquisitions. The time resolution is limited to 0.2 ms by  the antenna ADC 5 kHz sampling.

The gw detector EXPLORER   \cite{longterm} 
is located in CERN (Geneva-CH) at about 430 meters above sea level.  Scintillators counters were installed at EXPLORER in 2002, using scrap equipment recovered after the LEP shutdown.
Above the cryostat there is a single layer of 11 scintillators for a total area of $9.9~m^2$.
Below the cryostat there are two layers of 4 counters each, with a total area of  $6.3~m^2$.
Each scintillator is seen by two photomultipliers (56AVP). The signals  from the anode and  from the last dynode  of each 
photomultiplier  are digitized to measure the total charge. No large saturation effects occur in the dynodes up to particle densities of the order of $2000 \frac{particles}{m^2}$. The typical trigger threshold during normal run is of the order of $5 \frac{particles}{m^2}$ in the lower detectors. 

The detector is  calibrated using cosmic ray muons as in NAUTILUS.  Corrections are applied to take into account the scintillators attenuation lengths and the conversion of photons inside the scintillators. The  cosmic ray and the antenna data acquisition systems are independent, as in NAUTILUS. Similarly, a GPS clock is used to synchronize the acquisitions.

In order to compare the shower particle densities measured by the scintillators and by the streamer tubes, we have installed two scintillators, equal to the ones used in the top layer of EXPLORER, above  the NAUTILUS cryostat. The scintillators measured  numbers of particles $+20\%$ larger  than streamer tubes for showers particle densities around $400~\frac{particles}{m^2}$.

Most of the high energy events are due to electromagnetic showers. The rate of electromagnetic air showers (EAS) is computed starting from the empirical relation due to G. Cocconi  \cite{cocco}

\be
H(\ge\Lambda)=k\Lambda^{-\lambda}~events/day
\label{rscia}
\ee

where $\Lambda$ is   the density  of secondaries in an EAS, measured in units of number of charged particles per square meter, $\lambda=1.32 + 0.038$ $ln(\Lambda)$
and $k=3.54\times10^4$.
This relation holds at sea level and in absence of absorbing material. The NAUTILUS antenna is located inside a building  with a very small amount  of  matter in the roof, while EXPLORER is in a normal building with concrete roof. Concrete has 50 MeV critical energy,  to be compared to a critical energy 88 MeV for air; therefore
we expect  in the EXPLORER detector an increase of the electromagnetic showers particle density due to the different critical energies and to its above sea level higher location.

\begin{figure}
\includegraphics[width=5.5in,height=3in]{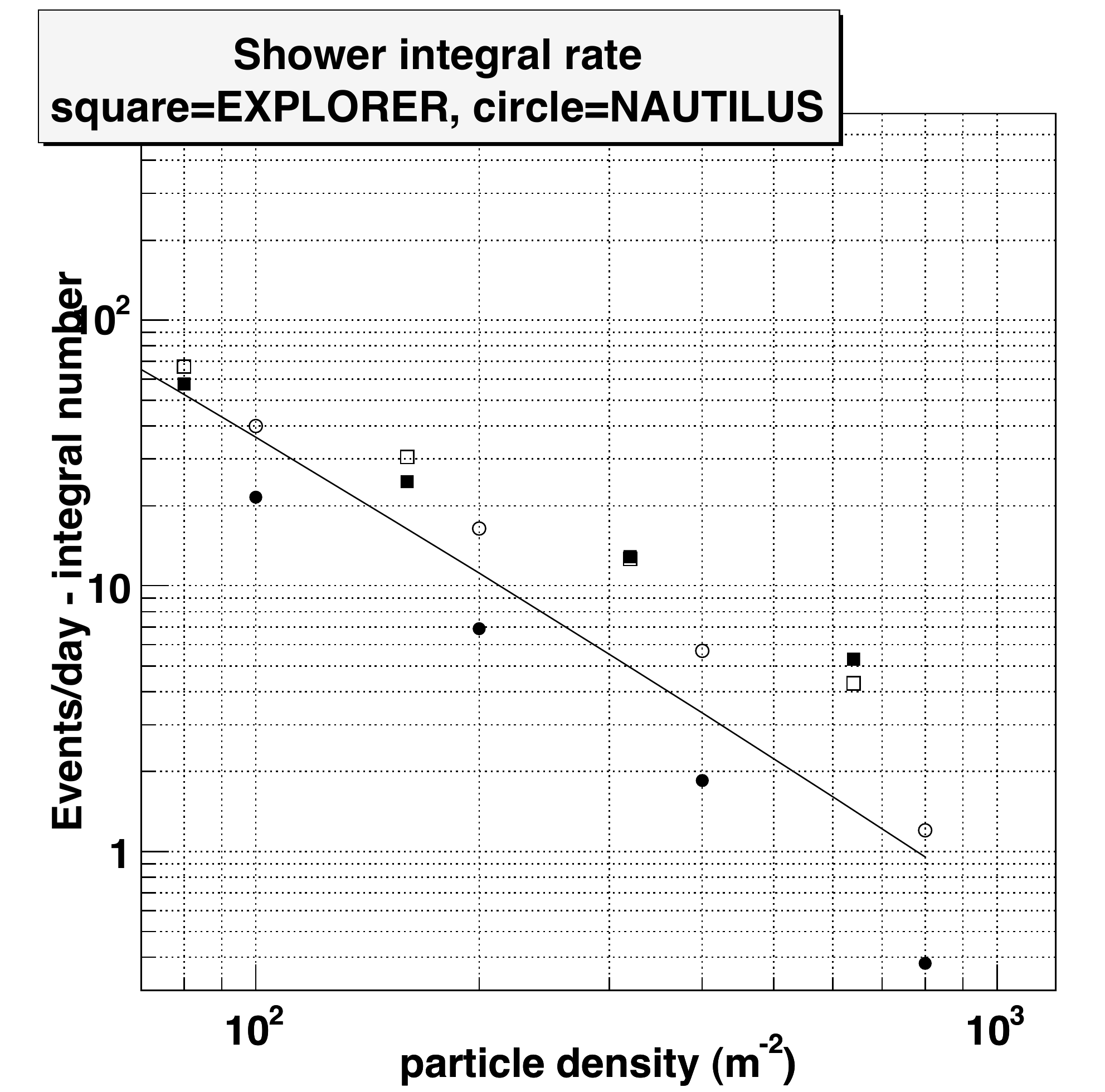}
\caption{The integral rate of events computed using Cocconi's relation (eq.\ref{rscia}) (continuous line), the rate measured by the EXPLORER  scintillators  (squares) and by the NAUTILUS streamer tubes (circles). The open symbols refer to the detectors below the cryostat, the filled symbols are for the detectors above. The EXPLORER data are corrected for the $20\%$ systematic difference in the number of particle measured with scintillators with respect to streamer tubes. The difference in the rates of the detectors above the cryostats is due to the different altitudes of Frascati and CERN and to the presence of a roof in the EXPLORER building. The rates in the bottom detectors are also influenced by the different distribution of iron, copper and aluminum in the cryostats.  
 \label{corate} }
\end{figure}

The  integral distribution of the showers measured with  both  EXPLORER and NAUTILUS  cosmic ray detectors is shown in  Fig.\ref{corate}.  In  this figure we have (arbitrarily) taken NAUTILUS  as reference: therefore we have shifted the EXPLORER points in order to correct  the systematic  $20\%$ error between streamer and scintillators. 

The NAUTILUS data of the detector above the cryostat  are in agreement with the  prediction of Eq.\ref{rscia} within the  $25\%$ systematic error given by the particle density measurement. The difference between NAUTILUS and EXPLORER  and between detectors over and under the cryostats are due to several, already mentioned effects:  the differences in altitude  of the experimental locations (230 m), the presence of a concrete roof in the EXPLORER building and the materials in the cryostats.

  The signal expected in a gw detector like  NAUTILUS,  a bar 3 m in length and 0.6 m diameter, as a consequence of  the interaction of a particle releasing an energy (W in GeV units) is \cite{cosmico1,cosmico2,cosmico3} , according to relation (\ref{eliub}):

\be
E\sim \frac{7.64}{2}\cdot10^{-9}W^2\alpha^2 \hskip1cm  [K[
\label{ww}
\ee
where the bar oscillation energy $E$ is expressed, as usual in the antenna jargon, in kelvin units ($1 K= 1.38\cdot 10^{-23} J$),  the numerical constant is the value computed using the linear expansion coefficient and the specific heat of pure aluminum at 4 K and $\alpha$  takes, as described in the previous section, either value  $ \alpha_{n}= 1.15$  above the $s$ transition  temperature or $ \alpha_{s}=3.7$   for superconductive Al 5056. The constant $7.64\cdot10^{-9}$ applies if the energy is released in the bar center. If the energy is uniformly distributed along the bar, as in the case of EAS showers,  this value is reduced by a factor 2.

The cosmic ray event rate in NAUTILUS has been evaluated considering three different event categories: pure electromagnetic showers, showers produced by muons and showers produced by hadrons in the bar. We use Eq.\ref{ww}  with the correction $\alpha_n=1.15$ for the response of an aluminum Al 5056 bar in the normal state. 

The rate of the EAS and the energy deposited  by an EAS has been computed starting from Eqs. \ref{rscia},\ref{ww} with the following assumptions:

 1) No particle absorbed (all particles go through the bar): indeed the radiation length in the bar is small compared to the total radiation length in the atmosphere. 
 
 2) The energy loss for a single particle is computed assuming ionization energy losses for electrons having the aluminum critical energy.
 
 3) We used the showers angular distribution as reported in\cite{Aglietta:1994ge}.
 
 4) We neglected the contribution of hadrons that could be present in the core of the showers.
 
 Under the previous assumptions and using  the density $\Lambda$ of secondaries we obtain
\cite{cosmico1,cosmico2,cosmico3}:

\be
E=\Lambda^2~4.7~10^{-10} \alpha^2  \hskip1cm  [K[
\label{theo}
\ee

The production of the showers due to muon and hadrons was computed using  the GEANT package\cite{geant}, developed at CERN, to simulate NAUTILUS  and the CORSIKA\cite{heck} Montecarlo, as input to GEANT, to simulate the effect of the hadrons
produced by the cosmic ray interactions in the atmosphere, assuming a cosmic ray "light" composition.
The Montecarlo simulation reflects 1 year of data taking.

 The results are  shown in Table \ref{table_1}.
 The rate of the the events scales as  $W^{-0.9}$. This is because 
 the cosmic ray integral spectrum is well described by a power law  $W^{-\beta}$ with  $\beta\sim1.7$  for cosmic ray primaries up to the so called "knee" at $W_{primary}=10^{15}eV$ and  $\beta\sim2$ at higher energies. The energy in the first longitudinal mode $E$ (first column of Table \ref{table_1}) is proportional to the square of the absorbed energy W.

\begin{table}[h]
\centering
\begin{tabular}{|c|c|c|c|c|c|c|c|}
\hline
Vibrational&Deposited&Muons&Ext Air &Hadrons&Total\\
Energy E & Energy W&& Showers&&\\
(K)&(GeV)&&&&(events/day)\\
\hline
$\ge10^{-5}$&$\ge44.5$&15.7&62&29.2 & 107  \\
$\ge10^{-4}$&$\ge141$&1.6 & 8.9 &4 & 14.5   \\
$\ge10^{-3}$&$\ge445$&0.2 &1 &0.4 & 1.6   \\
$\ge10^{-2}$&$\ge1410$&0.003 & 0.13 &0.06 & 0.19   \\
$\ge10^{-1}$&$\ge4450$& &  & & 0.03  \\
\hline
\end{tabular}
\caption{ Estimated rate (events/day) of antenna excitations due to cosmic rays in NAUTILUS  as a function of
the vibrational energy of the longitudinal fundamental mode that such events can produce.  The value at $E =0.1 K$ is obtained  extrapolating  from the lowest energy  values.
The values in the second column are the energies absorbed by the bar computed from  Eq. \ref{theo}, with the assumption of energy uniformly distributed,  and $\alpha_n=1.15$.}
\label{table_1}
\end{table}

There is quite a large uncertainty in the estimation of the high energy
event rate. This is due to uncertainties both in the cosmic ray composition and in the models of hadronic interactions at high energies. We have performed a check of the hadron flux at sea level used in our simulation with the direct hadron flux measured by the EAS-TOP experiment\cite{Aglietta:2002my}, properly scaling their results for the different altitude.
We have found  that at 1 TeV the EAS-TOP measurement gives a flux  roughly  $+35\%$ higher that the one used in our simulation.
This gives an idea of the uncertainty in the simulation of the hadronic effect.

Another uncertainty is due to  the EAS rate, modified by the presence of materials, as shown in Fig.\ref{corate}. Comparing the rates of the EXPLORER and NAUTILUS lower detectors and extrapolating  Fig.\ref{corate} at higher densities we have estimated that EXPLORER should have an excess of events respect to NAUTILUS of a factor 2.8  for energies larger than 0.1 K. We underline that, due to the large uncertainties involved, the expected absolute rate of events producing signals in a gravitational wave bar has also a large uncertainty,  of the order of that shown in Fig.\ref{corate}.  These, however affect in the same manner both our antennas, so that  the uncertainty on the relative rates of EXPLORER and NAUTILUS  is much smaller, being only due to systematic errors in the calibration of the EAS detectors ($\sim25\%$) and of the gravitational wave detectors  ($\sim10\%$).
In the following, we shall use only the particle densities measured by the lower detectors as they are closer to the bar.

\begin{figure}
\includegraphics[width=6in,height=3.5in]{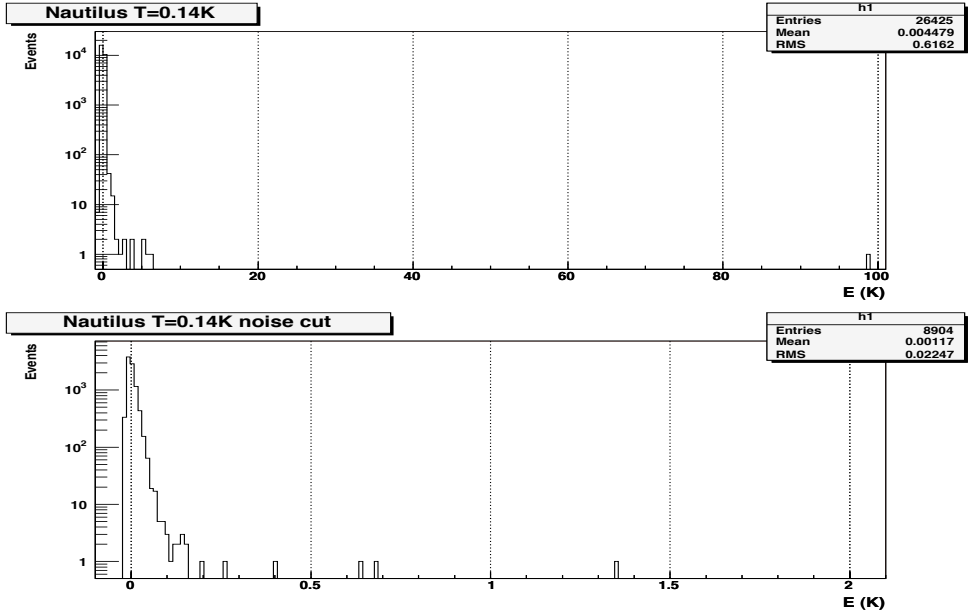}
\caption{NAUTILUS 1998. Distribution of the measured $E_{exp}$, with $\Lambda\ge 50~\frac{particles}{m^2}$. The upper graph shows all data. In the  lower graph a cut for $T_{eff}\le 5~mK$ was applied. The negative values are due to the  subtraction of the noise energy ($T_{eff}$ in the antenna jargon).
    \label{distri98} }
\end{figure}

\section{Antenna Signals Generated by Cosmic Rays}\label{antennas}
\subsection{NAUTILUS in 1998}

The ultra-cryogenic resonant-mass gravitational wave (gw) detector 
NAUTILUS \cite{rog} operating since 1996 at the INFN Frascati 
Laboratory, consists of a 3 m  2300 kg  Al 5056  alloy  bar.
The cryostat mainly  consists of seven concentric layers: three steel vessels, two thin aluminum plus  three thick copper  thermal shields.
During the run of 1998 it was cooled 
at 140 mK. The quantity that is observed (the "gw antenna output") is the vibrational amplitude of its first longitudinal mode of oscillation. This is converted by 
means of an electromechanical resonant transducer into 
an electrical signal which is amplified by a dc-SQUID. 
The bar and the resonant transducer form a coupled 
oscillator system, with two resonant modes, whose 
frequencies were, in 1998 $f_-= 906.40~ Hz$ and $f_+ =921.95~ Hz$.

The data regarding the vibrational energy of the NAUTILUS gw antenna 
were recorded with a sampling time of 
4.54 ms and processed with the delta-matched filter \cite{fast} optimized to detect impulsive signals. 
In a previous paper \cite{cosmico1} we reported the results of a search for correlations between the NAUTILUS data and the data of the EAS detector, when for the first time acoustic signals generated by EAS  were measured. In a further investigation \cite{cosmico2}, we found very large NAUTILUS signals at a rate much greater than expected. Now we know that, since the bar temperature was about 0.14 K,   the value $\alpha_s=3.7$ must be used in Eq. \ref{theo} to compute the expected response.

 \begin{figure}
\includegraphics[width=6in,height=3.5in]{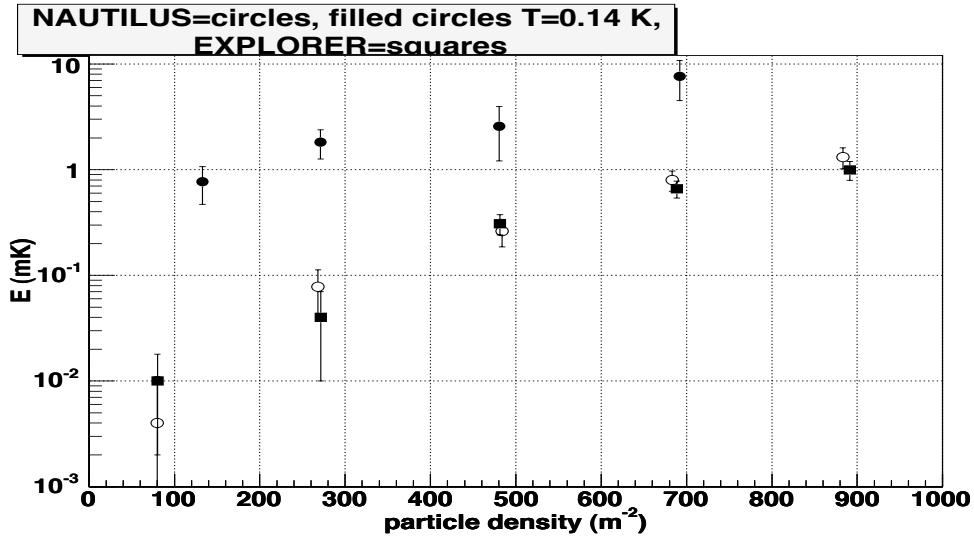}
\caption{Averages of signals with energy $E_{exp}\le0.1~K$, grouping data in ranges of particle density $\Lambda$. Filled circles NAUTILUS  at T = 0.14 K, open circles  NAUTILUS at $T = 3~K$, filled squares  EXPLORER  at $T=3~K$.
The data gathered at $T = 0.14~K$ are almost one order of magnitude larger than those collected at $T = 3~K$.
 \label{resp} }
\end{figure}

\begin{figure}
\includegraphics[width=6in,height=3.5in]{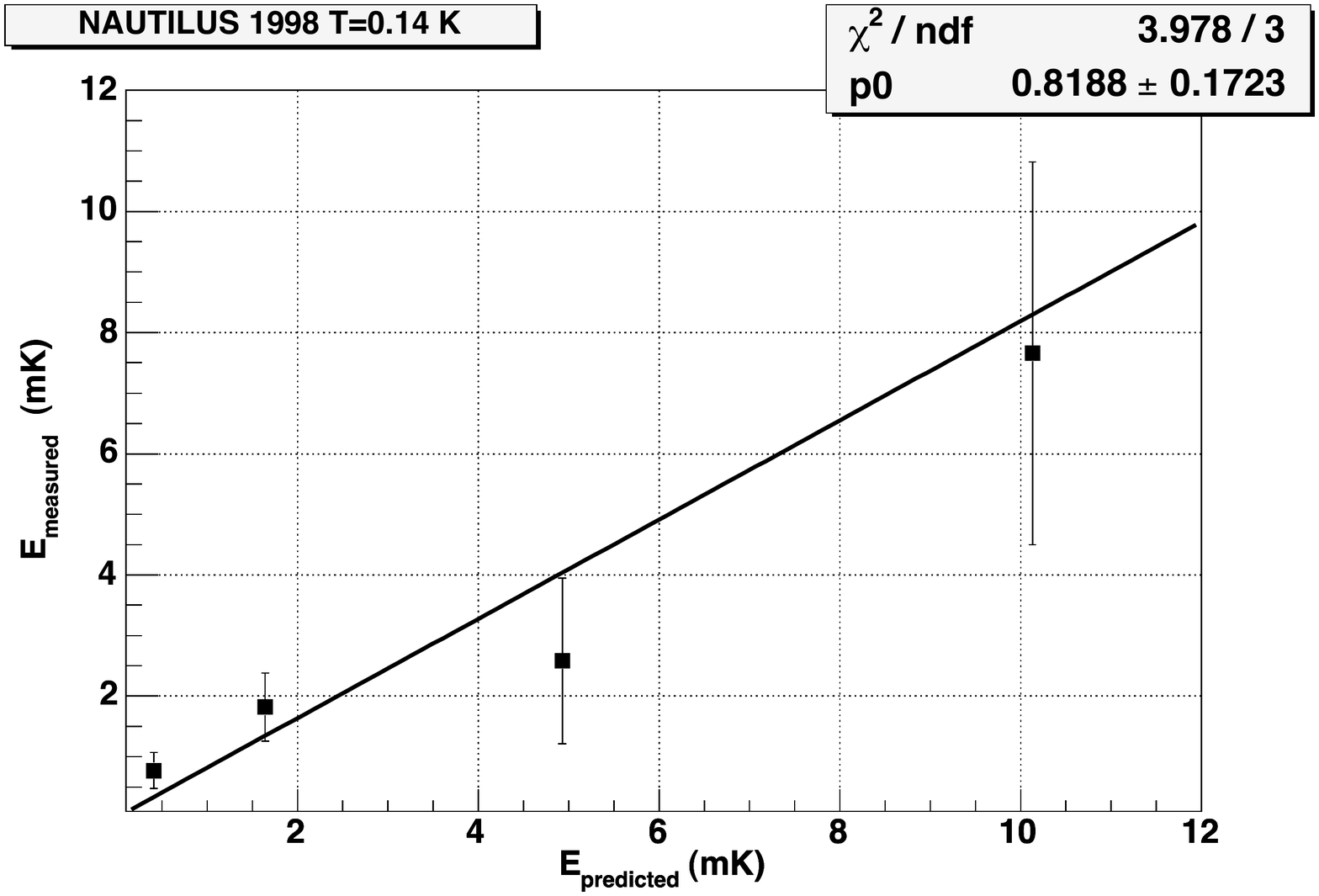}
\caption{NAUTILUS 1998: small signal analysis. Signals with energy $E_{exp}\le0.1~K$. Measured average energy $<E_{exp}>$ of signals with their standard deviations in K units versus signal energy as calculated with Eq. \ref{theo}, using the particle densities  measured by the streamer tubes below the cryostat and with $\alpha_s=3.7$.  The straight line is a least square fit through the origin, the slope is $p0=0.82 \pm 0.17$. The fit is in good agreement with the predictions, given the systematic errors on the number of particles ($\sim50\%$ ) and on the  antenna calibration ($\sim10\%$)
 \label{naut98e} }
\end{figure}

The correlation between the small signals detected by NAUTILUS and the impinging EAS has been described in detail in \cite{cosmico2}. The main points of this procedure are:
\begin{enumerate}
\item
We consider stretches of the filtered NAUTILUS  antenna data corresponding to EAS with
density $\Lambda$ , with a lower  threshold $\Lambda\ge 50\frac{particles}{m^2}$. 
 \item
For each stretch  we calculate the average energy $\bar{E}$,
in a time interval $ \pm 227~ms$ around the EAS arrival time, subtracting the value due to the noise energy ($T_{eff}$ in the antenna jargon). The time interval is chosen to take into account  the expected shape of the offline filtered signal.
 \item
 With this averaging procedure we avoid the problem of taking
either a maximum  or a minimum value, which may be due to noise and, when due to signals,
might not be exactly in phase among the various stretches.
By doing so we get average values $\bar{E}$.  In order to convert the value $\bar{E}$ into the energy at the maximum $E_{exp}$, we multiply $\bar{E}$ by a factor 4.1, as found, with a statistical dispersion of a few percent,  by numerically averaging the data sample of big events where the signal is much larger than the background, so that noise effects can be neglected.

\item
We obtain 26425 stretches of filtered data in coincidence with EAS. The NAUTILUS average noise level during this run was $T_{eff}\sim~10~mK$, while in the NAUTILUS 2003-2006 run the noise was $T_{eff}\sim~4~mK$.  In order  to perform a more meaningful comparison of these data with the NAUTILUS 2003-2006 run we have considered only those stretches with $T_{eff}\le 5~mK$.  In this way the number of useful stretches of filtered data reduces to 8904, in the period October 1998 to December 1998, for a total live-time of 27.3 days.
 \end{enumerate}

In order to verify the TAM model, we eliminate large  signals with energy  $E_{exp}\ge100~mK$ and we bin the remaining in  five ranges according to the particle density $\Lambda$,  measured by the streamer tubes under the cryostat with an upper cut to  $\Lambda=~1000 \frac{particles}{m^2}$ to avoid the saturation effects in the cosmic ray detectors.   

The plot of excitation energy $E_{exp}$ vs particle density $ \Lambda$ is shown in Fig.\ref{resp}. In this figure we show both  the measurements with NAUTILUS at 140mK and 2.6K, as well as EXPLORER at a temperature of about 3 K (see discussion in the following sections). We clearly see a difference of almost an order of magnitude between the measurements taken with  aluminum  in the $(s)$ state and those in  $(n)$ conduction state.

We apply Eq.\ref{theo} to evaluate the expected response according to the TAM model with using $\alpha_s=3.7$ . The results are shown in Fig.\ref{naut98e}: the agreement between the measurements and the values calculated with the TAM %thermo-acoustic model
 is reasonably good. The slope of the linear fit passing through the origin is $0.82\pm0.17$ compatible with one, and this confirms that the model works well also in the $(s)$ state. The systematic uncertainty in the number of particles is $\sim25\%$, therefore, as E scales quadratically with $\Lambda$  (see Eq. \ref{theo}) we expect an error in the determination of the slope of $\sim50\%$, plus a smaller ($\sim10\%$) uncertainty due to the antenna calibration . 
 
\begin{figure}
\includegraphics[width=6in,height=3.5in]{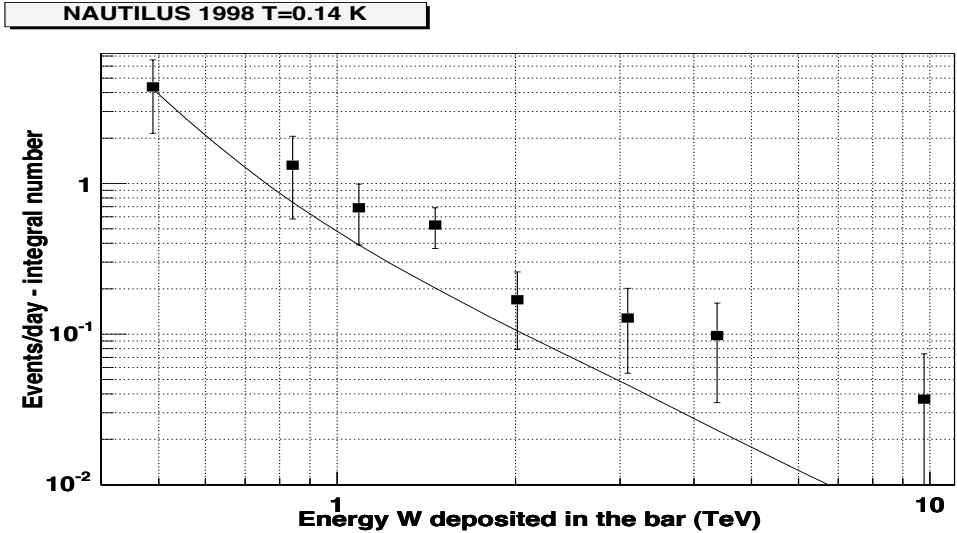}
\caption{ NAUTILUS 1998. The integral distribution of the event rate after the background unfolding,  compared with the expected distribution (continuous line). The prediction is computed using the data of Table \ref{table_1} 
and using the appropriate value $\alpha_s=3.7$.}
 \label{naut98a}
\end{figure}

We have also measured the rate of events producing signals in the bar. To this purpose we have used the energy distribution of Fig.\ref{distri98} with noise   $T_{eff}\le 5~mK$. This energy distribution is the convolution of the cosmic ray signals with the noise due to random fluctuations of the background.
We have computed the background energy distribution by shifting the time of the cosmic rays (20 different values  with 2 s  intervals for each cosmic ray).
The event rate per day after the unfolding of  the background distribution is shown in   Fig.\ref{naut98a}.  
The agreement with the predictions, computed from the figures of Table \ref{table_1} modified by using the correct value of $\alpha_s$,  appears very good (taking into account the very large uncertainties in the expected rates).

\subsection{NAUTILUS in 2003-2006}

In 2003 some components in the readout of NAUTILUS antenna were upgraded, similarly to what has been done in EXPLORER the previous year
\cite{Astone2003eq}.
  This resulted in an  increased bandwidth from the 1998 value of $0.4~Hz$  to about $9.6~Hz$. The data were since recorded with a sampling time of  0.2 ms and  processed with the delta-matched filter \cite{fast} . 
NAUTILUS  is operated at about 3 K by cooling with a superfluid  $\ ^{4}He$ bath, so that $\alpha_n=1.15$  is used in  Eq.\ref{theo} to estimate the signal expected according to the TAM.

\begin{figure}
\includegraphics[width=5.5in,height=3.5in]{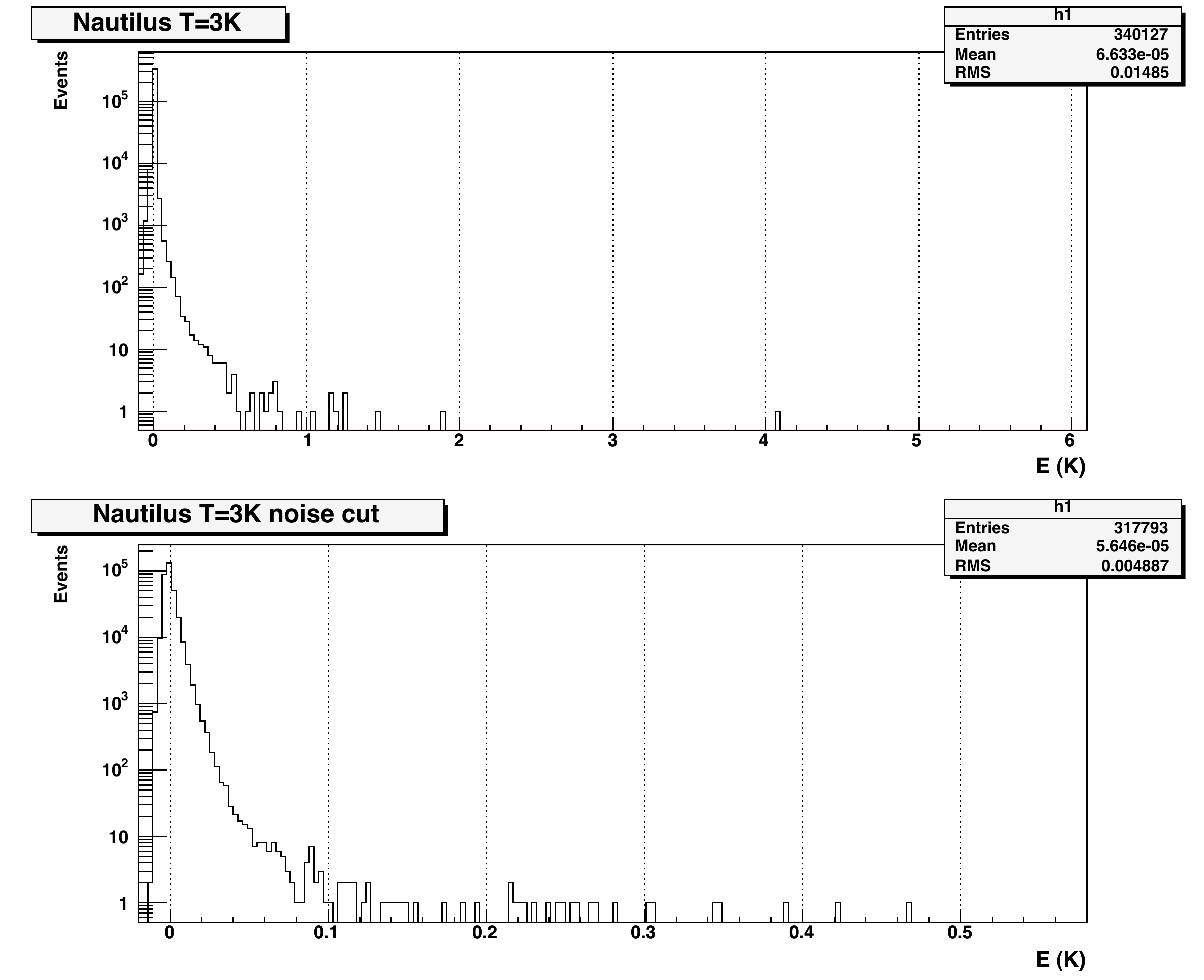}
\caption{NAUTILUS in 2003-2006. Distribution of the measured antenna output  $E_{exp}$, with $\Lambda\ge 50~\frac{particles}{m^2}$. The upper graph reports all data. The lower graph is for $T_{eff}\le 5~mK$.
    \label{distrituttina} }
\end{figure} 

\begin{figure}
\includegraphics[width=5.5in,height=3.5in]{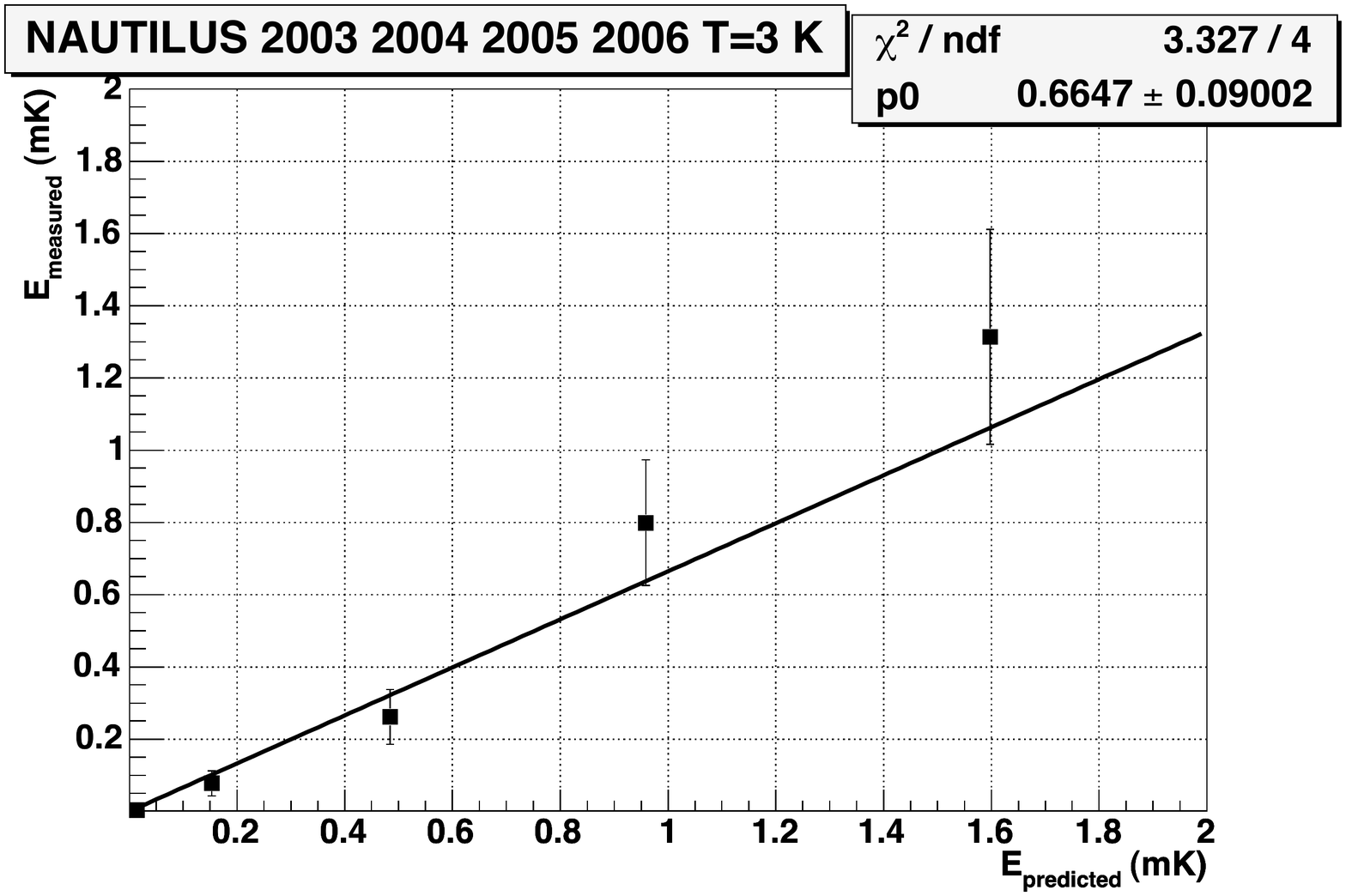}
\caption{NAUTILUS  2003-2006: small signal analysis. Signals with energy $E_{exp}\le0.1~K$ with their standard deviations compared to the signals calculated with Eq. \ref{theo} and $\alpha_n$  using the particle densities  measured by the streamer tubes under the cryostat. The straight line is a least square fit through the origin.  The fit is in good agreement with the prediction taking into account the systematic errors from  the measurement of the number of particles ($\sim50\%$ ) and from the antenna calibration ( $\sim10\%$)
 \label{nauttutti} }
\end{figure}

We have carried out the small signals analysis  according the following procedure,  similar to that outlined in of the previous section:

\begin{enumerate}
\item
The data for the four years 2003-2006  were grouped together  with a lower threshold $\Lambda\ge 50~\frac{particles}{m^2}$. 
 \item
For each stretch of filtered data the average energy $\bar{E}$ is calculated using 160 contiguous samples, corresponding to  $\pm 16~ms$ around the EAS arrival time. 
 \item
The adjusting factor from the averages $\bar{E}$ to  the maximum $E_{exp}$ is  2.3, as found by numerically averaging large events where noise contribution is negligible. The difference with respect to the previous value 4.1 is mainly due to the larger bandwidth of the detector and to the different time window.

\item
We obtain in total 88263 stretches of filtered data, corresponding to EAS, and  82330 with $T_{eff}\le 5~mK$.  The live-time of the data  after the cut $T_{eff}\le 5~mK$  is 1086 days. The distribution of $E_{exp}$ is shown in Fig.\ref{distrituttina}
 \end{enumerate}

 With the same procedure used for the 1998 data,  we consider 5 ranges according the multiplicity $\Lambda\ge50~\frac{particles}{m^2}$ and apply Eq.\ref{theo} to compute the expected values according to the TAM. %thermo-acoustic model. 
 The results are shown   in Fig.\ref{nauttutti}.
The ranges in this figure are smaller than the ones in Fig.\ref {naut98e}, as a consequence of the different values of $\alpha$ in superconductive and in normal state.
The slope of a linear fit through the origin is $0.66\pm0.09$. This value should be compared with $0.64\pm0.12$, that we obtain by dividing for $\alpha_n^2$ the published value $0.85\pm0.16$\cite{cosmico2}. 

The analysis of ref. \cite{cosmico2} only concerned  the data  of NAUTILUS 2001:  at that time NAUTILUS was still working in the narrow band mode with the old readout system. The excellent agreement demonstrates the stability of the apparatus and the accuracy of calibrations, even in largely different set-ups. Moreover we stress the  good agreement between measured and calculated quantities, within  the TAM.

\begin{figure}
\includegraphics[width=5.5in,height=3.5in]{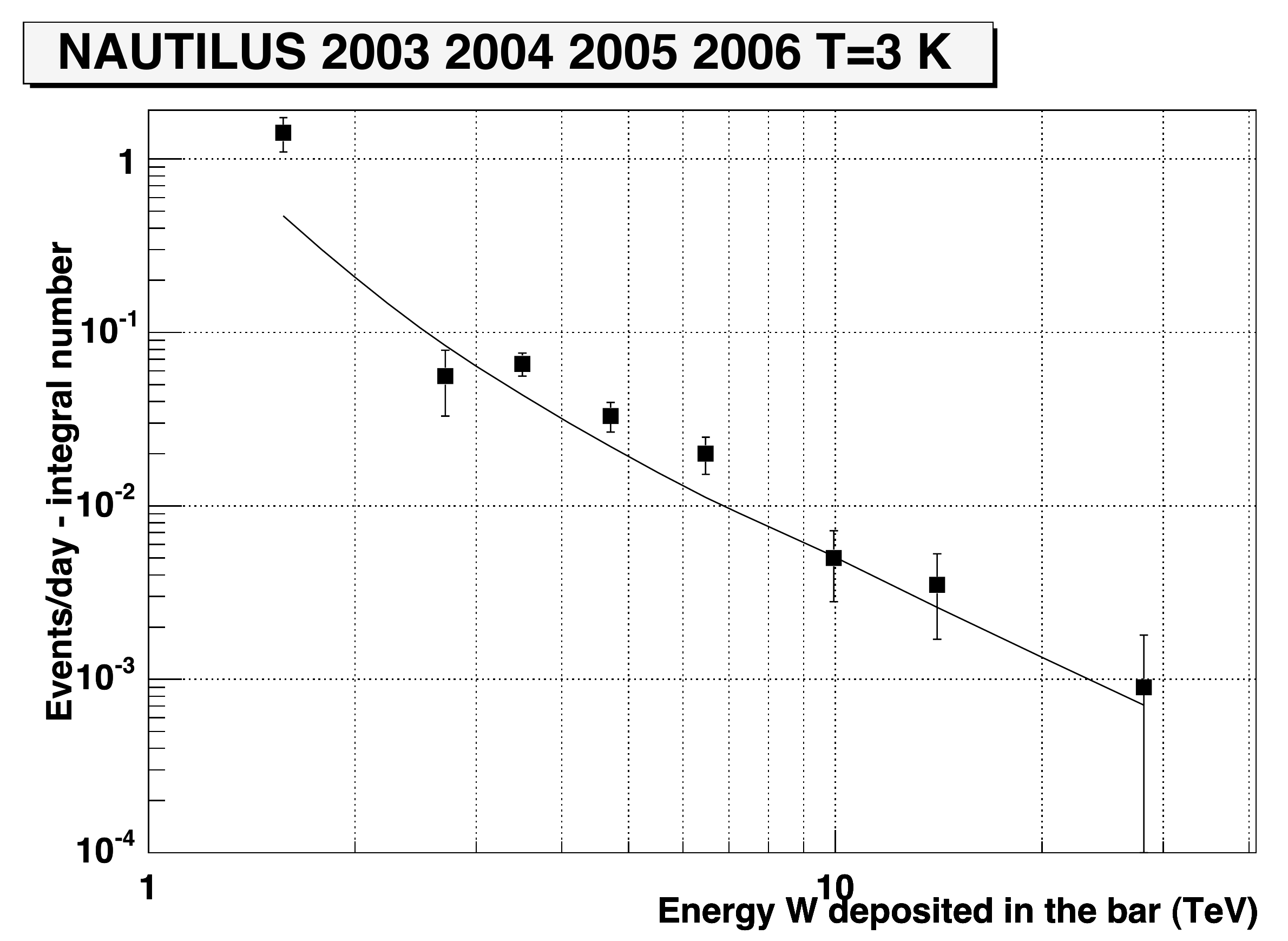}
\caption{ NAUTILUS   2003-2006 : The integral distribution of the event rate  after the background unfolding, as in fig.\ref{naut98a}, for the four years 2003-2006, compared with the expected distribution  (continuous line). The prediction is computed using  the data of Table \ref{table_1}.  
    \label{nauttuttie} }
\end{figure}

Figure \ref{nauttuttie} shows the integral distribution of the event rate after the background unfolding. 
The agreement with the expected distribution computed from Table~\ref{table_1}  is again quite good, considering the large uncertainties in the expected rates. The largest event detected in NAUTILUS in the years 2003-2006 has an energy of $E_{exp}\sim4.1~K$ in the first longitudinal mode corresponding to $W \sim 28~TeV$ in the bar.

\begin{figure}
\includegraphics[width=5.5in,height=3.5in]{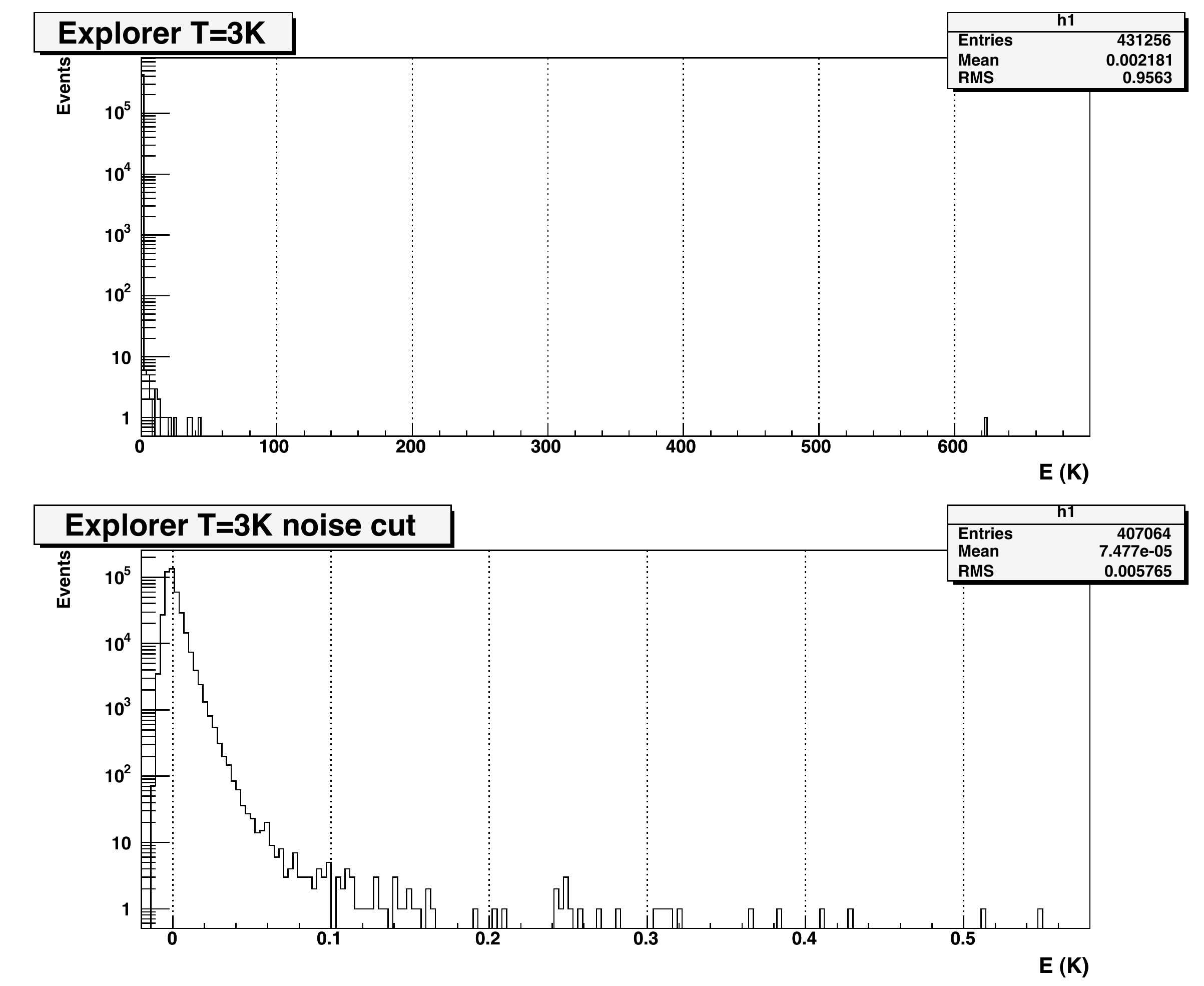}
\caption{EXPLORER in 2003 2004 2005 2006. Distribution of the measured antenna output $E_{exp}$, with $\Lambda\ge 50~\frac{particles}{m^2}$. The upper graph for all data. The lower graph for $T_{eff}\le 5~mK$. Note the very large event  with energy bigger than 600 K in the upper graph.
    \label{distrituttiex} }
\end{figure} 

\begin{figure}
\includegraphics[width=5.5in,height=3.5in]{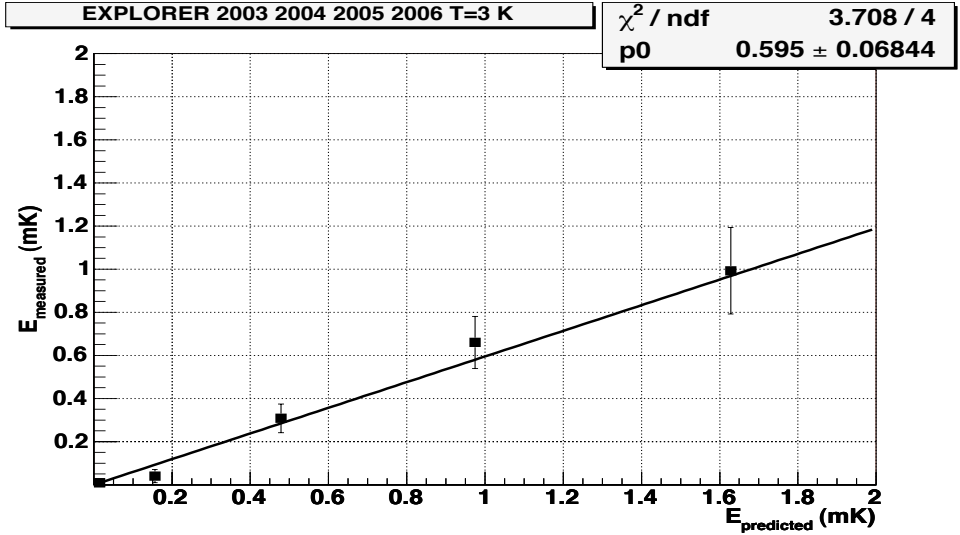}
\caption{EXPLORER 2003-2006: small signal analysis. Signal energy of events with $E_{exp}\le0.1~K$ with their standard deviations vs the energy calculated with Eq. \ref{theo} and $\alpha_n=1.15$,  using the particle densities  measured by the scintillators under the cryostat. The straight line is a least square fit through the origin.  The fit is in good agreement with the prediction taking into account the systematic errors from  the measurement of the number of particles ($\sim50\%$) and from the antenna calibration ($\sim10\%$). There is also a good agreement with  NAUTILUS data.
 \label{expttutti} }
\end{figure}

\begin{figure}
\includegraphics[width=5.5in,height=3.5in]{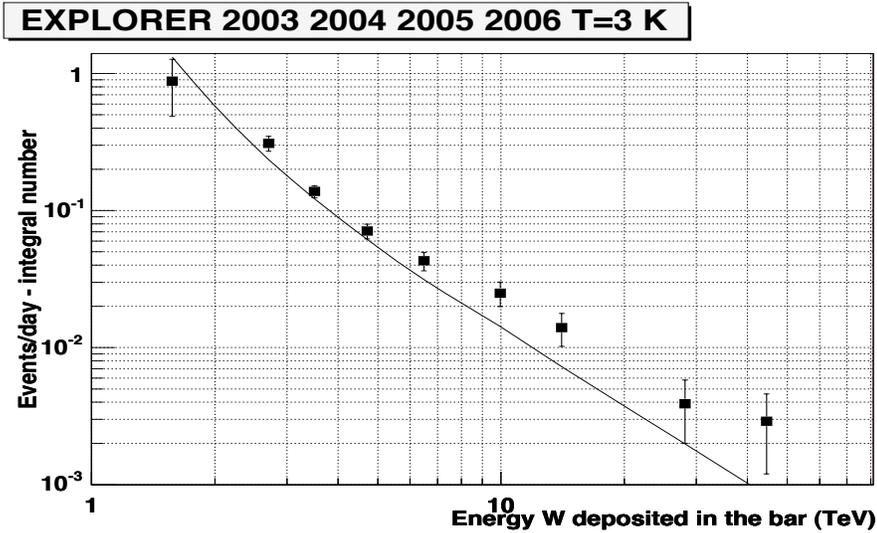}
\caption{ EXPLORER  2003-2006 :  The integral distribution of the event rate  after the background unfolding, as in Fig.\ref{naut98a},  compared with the expected distribution  (continuous line). The prediction is computed using Table \ref{table_1} multiplied by a factor 2.8 (see text).      \label{exptuttie} }
\end{figure}

\begin{figure}
\includegraphics[width=5.5in,height=3.5in]{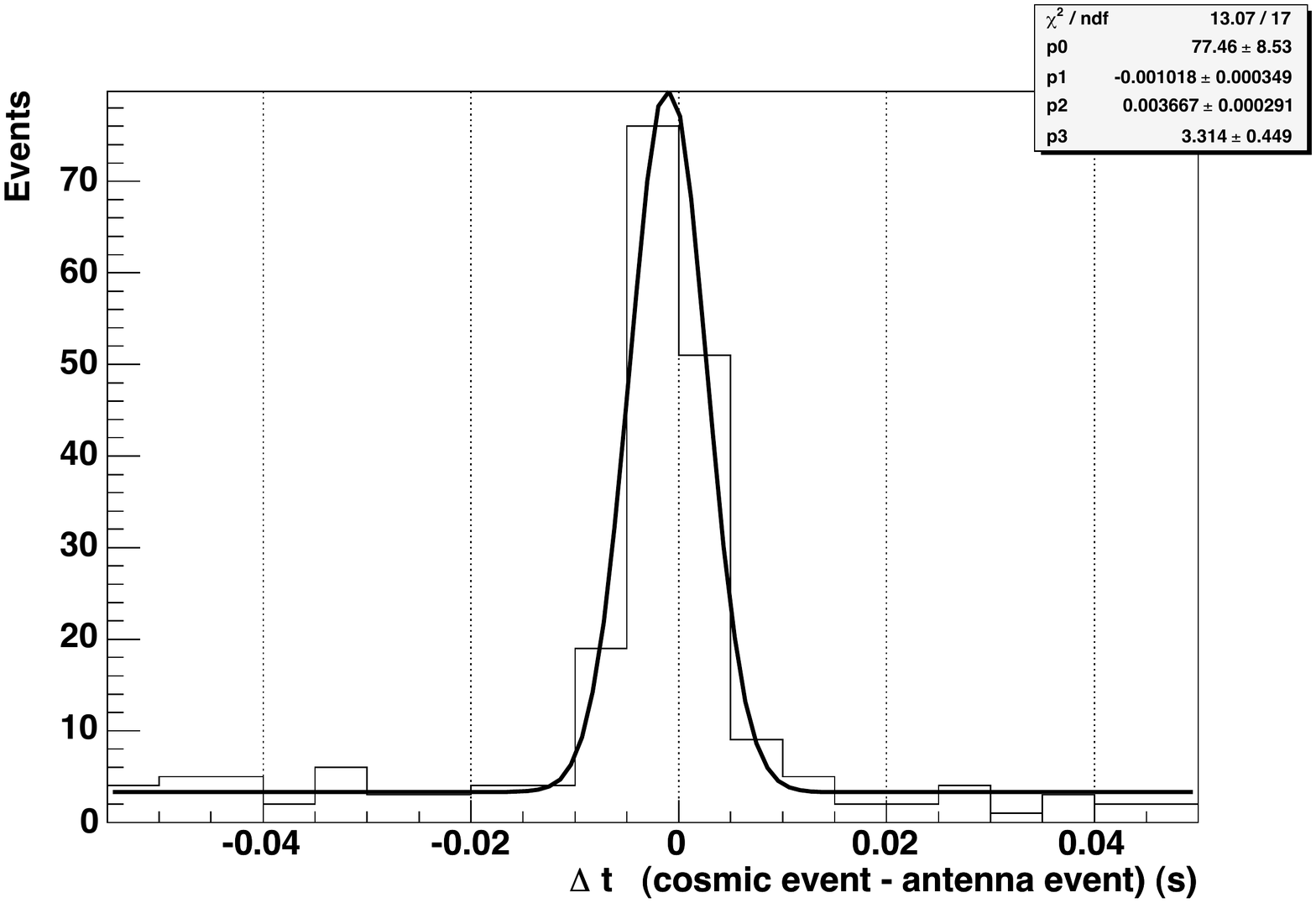}
\caption{ EXPLORER  2003-2006 : Time difference (seconds)  between  cosmic rays  with $\Lambda\ge 100~\frac{particles}{m^2}$ and the maximum of the filtered antenna signal, with a cut 
$E\ge 36~T_{eff}$. The fit with a gaussian , with parameters p0=peak, p1=mean, p2=$\sigma$ and a constant background p3, gives $\sigma= 3.7ms$. The value of the mean (-1$\pm0.35$ ms) should be compared to the expected value of -0.6 ms due to the delay of the antenna electronic chain.
    \label{ristempi} }
\end{figure}

\subsection{EXPLORER in 2003-2006}

The EXPLORER detector has been in almost continuous operation at CERN since 1991, and it has undergone over the years several upgrades that  progressively improved both its sensitivity and its operation duty cycle. 
EXPLORER has a bar similar to NAUTILUS, while the cryostat is sligthly different (three steel containers, one aluminum shield and a thin copper vessel). Data acquisition, readout and operation are very similar to NAUTILUS.  The operating temperature is $T\sim~2.6~K$. A detailed description of the apparatus and its main features (including data taking and analysis) can be found in ref \cite{longterm}. 
In 2001 EXPLORER   has been upgraded with a new read-out allowing for the first time "wide band" operation of a gw
bar detector\cite{Astone2003eq}.
Explorer operation was suspended in August 2002 due to a cryogenic failure. We took advantage of this stop to recondition the transducer and  complete installation of the cosmic ray shower detector described in section \ref{cosm_descr}.

For the small signal analysis we have applied the same procedure used for NAUTILUS, with the following differences: 

\begin{itemize}

\item
The adjusting factor from the averages $\bar{E}$ to  the maximum $E_{exp}$ is  2.5, again determined by numerically averaging large events. 
\item
We obtain in total 431256 stretches of filtered data, corresponding to EAS, and  407064  with $T_{eff}\le 5~mK$.  The live-time of the data  after the cut $T_{eff}\le 5~mK$  is 1022 days. The distribution of $E_{exp}$ is shown in Fig.\ref{distrituttiex}.
\end{itemize}

 Fig.\ref{expttutti} shows the small signal response analysis: as we did for  NAUTILUS, we plot the predicted energy vs. the measured energies (both expressed  in mK). The linear fit for the EXPLORER data  is compatible with that in  Fig.\ref{nauttutti} for NAUTILUS. This shows that we have in good control both the calibration of the cosmic ray detectors and the calibration of the gravitational wave detectors.

We also repeated the large event analysis and measured the rate of the large events. The results are in Figure \ref{exptuttie}, similar to Fig.\ref{distri98} and \ref{nauttuttie}.  The event rate in EXPLORER is higher that in NAUTILUS.  We expect a higher rate  due  the different altitude of Frascati and CERN and to the effect of the roof in the CERN building.
The continuous line in Fig.\ref{exptuttie} shows the predictions computed from Table~\ref{table_1} scaled by a factor 2.8 that accounts for the difference in the EAS rates as measured by the cosmic ray detectors and discussed in section \ref{cosm_descr} .

The agreement between measurement and expectations is again quite good, considering the large uncertainties in the calculation of the predicted rates.
It is important to note that acoustic gw detectors  have no large signal limitations  due to saturation effects and can detect very high energy events.

 Indeed the largest event detected up to now has an energy in the first longitudinal mode of $\sim670$ K corresponding to $\sim360$ TeV in the bar. The event occurred in EXPLORER on Nov 10 2006 9:40 UT.

\section{Conclusions}
We have discussed the NAUTILUS and EXPLORER response to small signals  (energy $E\le$ a few mK) and to large signals 
 (from energy   $E\ge~20~mK$ up to events of $E\sim~600K$).
 Table \ref{table2} shows the main results obtained. 
 
 \begin{table} [h]
\centering
\begin{tabular}{|c|c|c|c|c|c|c|}
\hline
RUN&T&small signal &expected&large ev rate  &large ev rate \\
        &(K)&slope &slope & $E>0.1K$& predicted\\
\hline
NA 1998      &$0.14$ &$0.82\pm0.17$&1$\pm0.5$          &$0.53\pm0.16$&0.24  \\
NA  2001\cite{cosmico3}      &$1.5$&$0.64\pm0.12$ &1$\pm0.5$            & $0.007\pm0.012$ &0.03  \\
NA 2003-06&$\sim3$&$0.66\pm0.09$      &1$\pm0.5$  &$0.034\pm0.007$&0.03 \\
EX 2003-06&$2.6$& $0.6\pm0.07$  & 1$\pm0.5$  & $0.078\pm0.01$ &0.08 \\
\hline
\end{tabular}
\caption{Data summary. The column "small signal slope" shows the results of the fits in Fig. \ref{naut98e}, \ref{nauttutti} and \ref{expttutti}. The predictions are based on RAP experiment \cite{rap}\cite{rap1}\cite{rap2}. The systematic error on the slope is $\sim50\%$. The errors reported in column three are statistical. The columns "large event rate" refer to the number of events per days having  energy $E>0.1K$.  
}
\label{table2}
\end{table}
 
 For small signals we have found that the acoustic gw  detector response well agrees with the predictions based on the 
 Thermo-Acoustic Model,  once the corrections to this model, provided by a dedicate experiment\cite{rap}\cite{rap1}\cite{rap2}, are applied. 
  
 The large signals  are the kind of events that can represent a background noise for current gravitational wave detectors. We found the rate of large signals to be in good agreement with the predictions, given the large uncertainty in such predictions. We notice that in EXPLORER we have a higher rate than in NAUTILUS, and this is due to the  the concrete roof and to the difference in altitude between Frascati and Geneva.
 
 We have shown that the unexpected large events detected in 1998 with NAUTILUS at T=0.14 Kelvin were due to  the superconductive state.

Cosmic rays noise could become an important noise in higher sensitivity detectors , namely in superconductive state, and this noise should be taken into account in possible future detectors of improved sensitivity, both acoustic \cite{Bonaldi:2006nj}\cite{deWaard:2003ug} and interferometric. As shown in this paper, cosmic rays can provide an useful tool to have a continuous monitor and calibration of the acoustic gravitational wave detectors. As an example Fig.\ref{ristempi} shows the antenna time resolution measured using the cosmic ray showers collected during the  EXPLORER run.

Very important is the fact that the observation of cosmic rays demonstrates that the detectors, both hardware and software, are indeed able to detect vibrations as small as $10^{-19}~m$.

\section{Acknowledgements}
We thank our technicians  F. Campolungo, G. Federici, M. Iannarelli,R. Lenci, R Simonetti , F. Tabacchioni, and E. Turri, for their help in building and running the detectors and for the cryogenic operations. 
This work is partially supported by the EU project ILIAS (RII3-CT-2004-506222).

\end{document}